\documentclass{elsart}
\usepackage{epsfig}
\usepackage{graphicx}
\usepackage{amsfonts}

\begin{document}

\begin{frontmatter}
\title{A simple model for magnetism in itinerant electron systems}
\author{Andre M. C. Souza}

\address{Departamento de Fisica, Universidade Federal de Sergipe,
 Sao Cristovao SE, 49100-000, Brazil}
\begin{abstract}
 A new lattice model of interacting electrons is presented. It can be viewed as a
 classical Hubbard model in which the energy associated to electron itinerance is
 proportional to the total number of possible electron jumps. Symmetry properties of
 the Hubbard model are preserved. In the half-filled band with strong interaction the
 model becomes the Ising model. The main features of the magnetic
 behavior of the model in the one-dimensional and mean-field cases are studied.
\end{abstract}

\end{frontmatter}

\section{Introduction}

The Heisenberg model\cite{Heis} expresses the energy dependence of localized
electron spins by means of the exchange interaction. This assumption creates
a starting point for an analysis on magnetic states in insulators. There are
a few well founded results for this model. For instance, one-dimensional
systems, the Bethe ansatz and bosonization solutions provide a set of exact
solutions of the model\cite{heis1d}. Rigorous results at zero temperature
have been obtained for two and three hypercubic lattices\cite{Liebheis}. By
numerical approach, physical properties are exact results for finite systems
and have been conjectured as an exact result for the properties of infinite
systems\cite{heisnum}. More generallly, the study of the Heisenberg model is
made using methods of approximated solutions. However, these approaches
certainly contain partial information that can differ from the exact
solution. Another natural possibility is to introduce a new model that makes
the analytical analysis easier or that can evidently be solved exactly. It
is important, however, that the model remains appropriate for describing
magnetic systems. This picture can be emphasized considering the Ising model%
\cite{Ising} which is the most important simplification of the Heisenberg
model. The crucial assumption is that the Ising Hamiltonian is explicitly
diagonal, but nontrivial from the point of view of statistical mechanics.

The Ising model has played a key role in the development of the theory of
interacting systems. It has a phase transition at finite temperature that
can be worked out with mathematical rigor and yields an exact solution in
two dimensions. Although it has only been considered a theoretical model for
many years, experiments have brougt forth evidence of its applicability for
real magnetic materials (see \cite{wolf} and references therein). It should
be noted that another virtue of the model is its correspondence with systems
such as lattice gases, binary alloys and neural networks among others\cite%
{huang}.

In conductors, the study of magnetism is more complex because the electrons
of magnetic states are often not localized. The basic model to describe the
interaction between electrons having translational degrees of freedom is the
Hubbard model\cite{Hub}.

Suggested in 1963, the Hubbard model also much like the Heisenberg model
remains with many of its basic features in debate. An important problem to
be answered is to establish which limits within the parameters of the
Hubbard model that favor ferromagnetism when neither the electronic
interaction nor the electronic itinerance favors this situation. Many
believe that ferromagnetissm can only be observed when there are more than
one electronic orbital per site. In the simple Hubbard model when only one
orbital per site is present the induced antiferromagnetism due to the
electronic itinerant term disfavors ferromagnetism. However, in the extreme
limits of electronic repulsion and in special lattice cases, we can find
ferromagnetism in this model\cite{Nagao}. Indeed, the small number of
accurate solutions (one-dimensional systems, ground state theorems and
numerical approach)\cite{LiebWu,Hirsh} and the difficulty of studying the
Hubbard model indicate that the understanding of this problem remains vague.
For example, we do not have an accurate solution for the transition metals.

Knowing the difficulties in dealing with the Hubbard model and the
importance of studying the effect of the electronic itinerance term, I will
shortly introduce a model that remains with the essential features of the
Hubbard model, but simplifies it using a classical situation for the
hopping. Despite the simplification, this model must be capable of
describing the magnetism of materials with itinerant electrons. This paper
is organized as follows. In Sec. \ref{sec:Model} the model is described. The
one-dimensional ring considering nearest-neighbor hopping is solved in Sec. %
\ref{sec:One-dimensional}. Sec. \ref{sec:Mean-field} presents the mean-field
solution for the model. Sec. \ref{sec:Conclusion} is devoted to conclusions.

\section{Model}

\label{sec:Model}The starting point for the present model is the Hubbard
model, which has the following Hamiltonian 
\begin{equation}
\aleph =-\sum_{ij\sigma }t_{ij}a_{i\sigma }^{\dag }a_{j\sigma
}+U\sum_{i}n_{i\uparrow }n_{i\downarrow },  \label{hami}
\end{equation}%
where $a_{i\sigma }^{\dag }$, $a_{i\sigma }$ are the creation and
annihilation operators for electrons of spin $\sigma $ at site $i$. The
density of electrons is denoted by $n_{\sigma }=\frac{1}{N}%
\sum_{i}n_{i\sigma }$ where $N$ is the number of lattice sites, and $%
n_{i\sigma }=a_{i\sigma }^{\dag }a_{i\sigma }$ is the number operator for
electrons of spin $\sigma $ at site $i$. The first term of the Hamiltonian
corresponds to the hopping of the electrons between sites $i$ and $j$, and $%
\ t_{ij}$ is the hopping integral representing the overlap of electron wave
functions. Usually, the overlapping is assumed only over nearest neighbour
sites. The second term represents the on-site Coulomb interaction $(U)$
between the electrons.

The crucial ingredient for the construction of the model is to consider the
energy associated to the electron itinerance as being equal to the total
number of possible electron jumps. Hence, it is easy to see that the
Hamiltonian can be defined as

\begin{equation}
\aleph =-\sum_{ij\sigma }t_{ij}n_{i\sigma }(1-n_{j\sigma
})+U\sum_{i}n_{i\uparrow }n_{i\downarrow }.  \label{hami2}
\end{equation}%
In this case, the hopping term commutes with the Coulombic term. The states
of the number operator are eigenstates of the Hamiltonian. As in the Ising
model, the Hamiltonian is explicitly diagonal and the model is much easier
to study compared to the Hubbard model.

The Hamiltonian of Eq. ($2$) can be seen as an approximation of the
extended Hubbard model, constructed by the inclusion of a third term in the
simple Hubbard model (Eq. $1$) which describes the off-site Coulomb
repulsion between electrons\cite{Emery},\cite{Hub2},\cite{exte}. The
off-site interaction has been appropriate for compounds that exibit
near-neighbors interaction compared to the bandwidth, in particular quasi
one-dimensional organic conductors\cite{Hub2}. The Hamiltonian of Eq. ($2$)
is the zero-band-width limit of the extended Hubbard model assuming off-site
interaction only between electrons of equal spins. This assumption can be
justified\ considering that the electron repulsion of different spins has
its main contribution on the same site, while electron repulsion of equal
spins contributes only to different sites by the Pauli exclusion principle.
Hubbard showed that the zero-band-width limit of the one-dimensional
extended Hubbard model can present an interpretation of the optical spectra
of many tetracyanoquinodimethane salts. Indeed, some features of finite
bandwidth may well survive in the present model if the interaction effects
are of dominant importance\cite{Hub2}. These considerations indicate to be
reasonable the application of the model represented by the Hamiltonian of
Eq. (2) to the  quasi one-dimensional organic conductors as well as to other
materials with the same electronic structure.

Since we wish a model capable of describing the magnetic properties
of materials with itinerant electrons, it should be important to observe
that relevant symmetry transformations of the Hubbard model are preserved,
such as the U(1) charge symmetry and the particle-hole transformation in the
case of bipartite lattice. Notice that this has an important consequence:
the condition for the chemical potential $\mu =U/2$ for the half-filled
band, independently of the temperature, is identical to that of Hubbard
model. It is also crucial to observe the ground state configurations of the
new model. Here, only general features will be presented. Detailed analyses
depend on specific works using this model. It is easy to see that at
half-filled band the two N\'{e}el antiferromagnetic configurations are
ground states of the new Hamiltonian for $U>0$ in non frustrated lattices.
When the number of electrons is smaller than half of the number of lattice
sites the ground state becomes highly degenerate and the ferromagnetic
configuration has also the lower energy. The model favors the
antiferromagnetic order, but for low or high electron densities there is a
tendency towards the coexistence of ferromagnetism with antiferromagnetism.
As will be observed in the one-dimensional case, for the half-filled band
the antiferromagnetic ordered phase exists only at zero temperature. With
finite temperature the ordered phase does not exist.

In the square lattice, for the half-filled band case, the ground state is
antiferromagnetic, but at finite temperature($T$) the solution has not yet
been found. For this same lattice, the Hubbard model is not solvable and
even its ground state is not known. Using the Monte-Carlo approach, Hirsch%
\cite{Hirsh} predicts antiferromagnetic ordered ground state only at $T=0$.
I believe that both the new model and the Hubbard model yield a similar
result to the relation between the Ising and Heisenberg models. With finite
temperature, both the new and the Ising models have an ordered phase while
the Hubbard and the Heisenberg models do not\cite{Mermin}.

\section{One-dimensional solution}

\label{sec:One-dimensional}One sees that in this case the Hamiltonian (2)
can be written as $\aleph =\sum_{i}E_{i}$, where 
\begin{equation}
E_{i}=t(n_{i\uparrow }n_{i+1\uparrow }+n_{i\downarrow }n_{i+1\downarrow
})+Un_{i\uparrow }n_{i\downarrow }-t(n_{i\uparrow }+n_{i\downarrow }).
\label{energia}
\end{equation}%
Let us take $t_{ij}=t>0$ over nearest neighbor sites, otherwise $t_{ij}=0$.
The thermodynamic properties of the model can be obtained using the transfer
matrix method. The grand-canonical partition function, at temperature $T$,
is written as $Z=\sum_{i}\lambda _{i}^{N}\ $, where $\lambda _{i}$ are
eigenvalues of the transfer matrix

\begin{equation}
\widehat{X}_{i},_{i+1}=e^{-\left( E_{i}-\frac{\mu }{2}\sum_{\sigma
}(n_{i\sigma }+n_{i+1\sigma })\right) /k_{B}T},
\end{equation}%
where rows and columns are labelled by $n_{i\sigma }$ and $n_{i+1\sigma }$,
respectively. The matrix $\widehat{X}$ is 4x4 and will be numerically
diagonalized . The largest eigenvalues $\lambda _{\max }$of $\widehat{X}$
are found. Considering the thermodynamic limit ($N\rightarrow \infty $), the
\ free energy is $A/N=-k_{B}T\ \ln \lambda _{\max }$, and the thermodynamic
properties follow from properly differentiating the free energy. For the
half-filled band the analytical expression for $\lambda _{\max }$ is

\begin{equation}
\lambda _{\max }=\frac{(1+y)(1+w)+\sqrt{(1+y^{2})(1+w)^{2}-2y(1-6w+w^{2})}}{%
2wy},
\end{equation}%
where $w=\exp [-2t/(k_{B}T)]$ and $y=\exp [-U/(2k_{B}T)]$. Figure 1 presents
the specific heat $C$ as a function of temperature for typical values of $%
U/t $ for the half-filled band. For $U/t\leq 13.5$ it has a peak. Whose
value increases with $U/t$ until the maximum $C/Nk_{B}=1.38$ for $U/t=0.7$
and decreases for larger values of $U/t$. For $U/t>13.5$ the peak splits in
two, which reflects a rearrangement of the fermionic structure in the system%
\cite{meu}. The picture presented is analogous to those ones of the
one-dimensional half-filled band Hubbard\cite{shiba} and Falikov-Kimball
models\cite{meu}. However, the critical values of $U/t$ which causes this
splitting are different: $U/t=0$ for the Falikov-Kimball model, $U/t=4$ for
the Hubbard model and $U/t=13.5$ for the present model. For $U/t\rightarrow
\infty $ the second peak (high-temperature peak) gets further from the first
peak as $T\rightarrow \infty ,$ and then the first peak yields the specific
heat with the maximum value of $C/Nk_{B}=0.44$ which is exactly the specific
heat of the Ising model\cite{bonner}. Figure 2 shows the temperature
dependence of the magnetic susceptibility $\chi $ for the several choices of 
$U/t$ in the half-filled case. The $\chi $ has a maximum and vanishes at
zero temperature. For large $U/t$ the magnetic susceptibility is well
described by the Ising model as was evidenced by the specific heat in Fig.
1. Finally, it is obtained without further difficulties no phase transition
at any finite $T$ and no long-range order for all $T>0$.

\begin{figure}
\includegraphics[width=80mm]{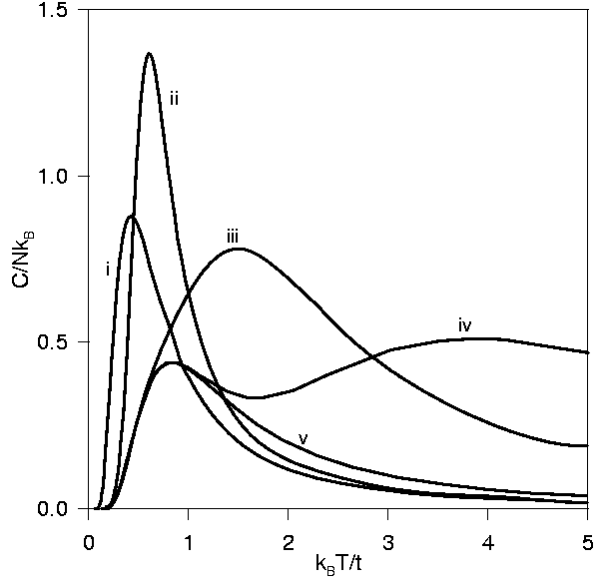}
\caption{Specific heat $C/Nk_{B}$ of
infinite chain versus temperature for typical values of $U/t$ : i) $U/t=0$;
ii) $U/t=1$; iii) $U/t=8$; iv) $U/t=20$; v) $U/t\rightarrow \infty $.}
\end{figure}

\begin{figure}
\includegraphics[width=80mm]{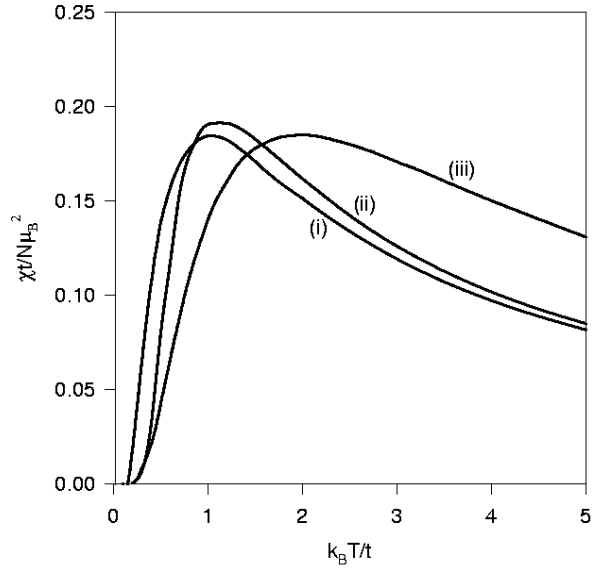}
\caption{Magnetic susceptibility of
infinite chain versus temperature for typical values of $U/t$ : i) $U/t=0$;
ii) $U/t=1$; iii) $U/t=32$.}
\end{figure}

\section{\protect\bigskip Mean-field solution}

\label{sec:Mean-field}Having obtained solutions for the one-dimensional case
one presents below the mean-field solution. The self-consistent equations
for translationally invariant systems are given by

\begin{equation}
n=\frac{2\cosh \left( \beta \left( \frac{mtz}{2}-h\right) \right) e^{\beta
\left( \mu +tz(1-n)/2\right) }+2e^{\beta \left( 2\mu -U\right) }e^{\beta
tz(1-n)}}{1+2\cosh \left( \beta \left( \frac{mtz}{2}-h\right) \right)
e^{\beta \left( \mu +tz(1-n)/2\right) }+e^{\beta \left( 2\mu -U\right)
}e^{\beta tz(1-n)}}
\end{equation}%
and

\begin{equation}
m=\frac{2\sinh \left( \beta \left( \frac{mtz}{2}-h\right) \right) e^{\beta
\left( \mu +tz(1-n)/2\right) }}{1+2\cosh \left( \beta \left( \frac{mtz}{2}%
-h\right) \right) e^{\beta \left( \mu +tz(1-n)/2\right) }+e^{\beta \left(
2\mu -U\right) }e^{\beta tz(1-n)}}
\end{equation}%
where $n$ and $m$ are the mean values of the electron number ($%
n=<n_{i}>=<\sum_{i\sigma }n_{i\sigma }>$) and local magnetization ($%
m=<m_{i}>=<\sum_{i}\left( n_{i\uparrow }-n_{i\downarrow }\right) >$),
respectively, on a lattice where each site has $z$ nearest neighbours. In
the half-filled band $n=1$, which implies that $\mu =U/2$ independently of
the temperature and it is easy to see that $m=\left[ \coth \left( \beta
\left( \frac{mtz}{2}-h\right) \right) +e^{-\beta U/2}\right] ^{-1}$. The
dependence of the magnetization on $U/t$ is considered. The critical
temperature $T_{c}$ is (assuming $k_{B}=1$, $zt=1$)

\begin{equation}
U=-2T_{c}\ln \left( \frac{1-2T_{c}}{2T_{c}}\right) .
\end{equation}%
The phase diagram is shown in Fig. 3. Below $T_{c}$ there is a spontaneous
magnetization. The inset in Fig. 3 shows the temperature dependence of the
magnetization. The magnetization increases with decreasing temperature and
attains the saturate value at zero temperature. Mean-field solutions suggest
that the\ magnetic order phase increases if $U/t$ increases in agreement
with the Hubbard model predictions\cite{Hirsh}.

\begin{figure}
\includegraphics[width=80mm]{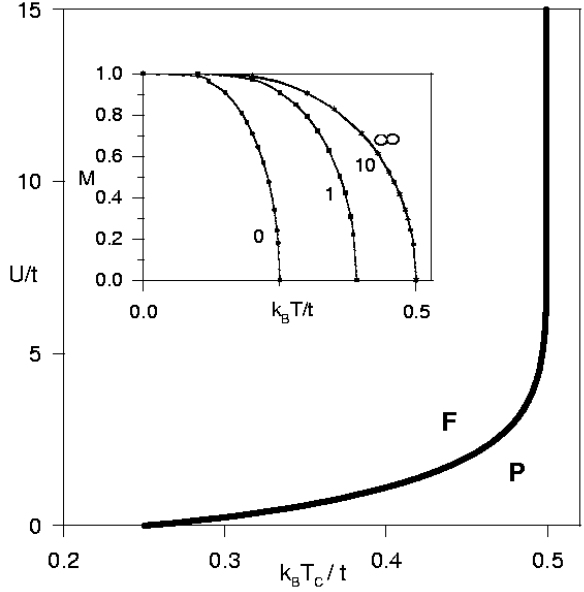}
\caption{Phase diagram for the
ferromagnetic (F) and paramagnetic(P) states of the model in the half-filled
band. The inset shows the temperature dependence of the magnetization. The
numbers denote the $U/t$ values.}
\end{figure}

\qquad\ 

Notice that for the $U=0$ limit the model is reduced to a lattice gas of two
types of atoms. This means that the present model can also be interpreted as
a generalization of the Ising model which is exactly the $U/t\rightarrow
\infty $ case.

\section{Conclusions}

\label{sec:Conclusion}In summary, a classical lattice model was introduced
to describe the magnetism of itinerant electron systems. It is much easier
to analyze than the Hubbard model. I have illustrated some features of its
magnetic properties. These general results clearly show that the presented
model can be viewed as a classical Hubbard model in the same way as the
Ising model can be considered with relation to the quantum Heisenberg model.
Naturally, it is not able to capture the quantum frustration of the Hubbard
model much like the Ising model cannot capture the quantum frustration of
the Heisenberg model. However, I believe that it is a promising help to
qualitatively clarify our understanding on the nature of magnetism. In
addition, the proposed model is a generalization of the Ising model. The
possibility of generalizing the Onsager solution for the square lattice case
would be useful in future phase transition research.

\section{\protect\bigskip Acknowledgements}

I thank M. E. de Souza and A. Reikdal for reading the manuscript. This work
was financially supported by CNPq and FAP-SE.


\begin{thebibliography}{99}
\bibitem{Heis} W. Heisenberg, Z. Phys. \textbf{49}, 619 (1928).

\bibitem{heis1d} H. Bethe, Z. Phys. \textbf{71}, 205 (1931); Y.-L. Liu.
Phys. Rev. Lett. \textbf{79}, 293 (1997); K. B\"{a}rwinkel, P. Hage, H.-J.
Schmidt, and J. Schnack. Phys. Rev. B \textbf{68}, 054422 (2003).

\bibitem{Liebheis} K. Harada, N. Kawashima, and M. Troyer. Phys. Rev. Lett. 
\textbf{90}, 117203 (2003); E. Manousakis, Rev. Mod. Phys. \textbf{63}, 1
(1991); T. Kennedy, E. H. Lieb, and S. Shastry, J. Stat. Phys. \textbf{53},
1019 (1988); K. Kubo and T. Kishi, Phys. Rev. Lett. \textbf{61}, 2585
(1988); F. J. Dyson, E. H. Lieb, and B. Simon, J. Stat. Phys. \textbf{18}, 335 (1978).

\bibitem{heisnum} J. C. Bonner and M. E. Fisher, Phys. Rev. \textbf{135},
A640 (1964); K. Lefmann, C. Rischel, Eur. Phys. J. B \textbf{21}, 313
(2001); A. A. Cuccoli, T. Roscilde, V. Tognetti, R. Vaia, and P. Verrucchi,
Phys. Rev. B \textbf{67}, 104414 (2003).

\bibitem{Ising} E. Ising, Z. Phys. \textbf{31}, 253 (1925).

\bibitem{wolf} W. P. Wolf. Braz. Journal of Phys. \textbf{30}, 794 (2000).

\bibitem{huang} K. Huang. Statistical Mechanics ( John Wiley and Sons, New
York, 1987).

\bibitem{Hub} J. Hubbard. Proc. R. Soc. London Ser. A \textbf{276}, 238
(1963).

\bibitem{Nagao} Y. Nagaoka, Phys. Rev. \textbf{147}, 392 (1966).

\bibitem{LiebWu} E. H. Lieb and F. Y. Wu., Phys. Rev. Lett. \textbf{20},
1445 (1968); R. Strack and D. Vollhardt, Phys. Rev. Lett. \textbf{72}, 2425
(1994); H. Tasaki, Phys. Rev. Lett. \textbf{69}, 1608 (1992); S. A. Trugman,
Phys. Rev. B \textbf{42}, 6612 (1990); E. H. Lieb, Phys. Rev. Lett. 
\textbf{62}, 1201 (1989).

\bibitem{Hirsh} J. E. Hirsch, Phys. Rev. B \textbf{31}, 4403 (1985).

\bibitem{Emery} V. J. Emery, Phys. Rev. B \textbf{14}, 2989 (1976).

\bibitem{Hub2} J. Hubbard, Phys. Rev. B \textbf{17}, 494 (1978).

\bibitem{exte} S.-J. Gu, S.-S. Deng, Y. -Q. Li, and H. -Q. Lin, Phys. Rev.
Lett. \textbf{93}, 086402 (2004); Y. Z. Zang, Phys. Rev. Lett. \textbf{92},
246404 (2004); S. Onari, R. Arita, K. Kuroki, H. Aoki, Phys. Rev. B 
\textbf{70}, 094523 (2004).

\bibitem{Mermin} N. D. Mermin and H. Wagner. Phys. Rev. Lett. \textbf{17}, 1133 (1966).

\bibitem{meu} C. A. Macedo, L. G. Azevedo, and A. M. C. Souza, Phys. Rev. B 
\textbf{64}, 184441 (2001).

\bibitem{shiba} H. Shiba and P. A. Pincus, Phys. Rev. B \textbf{5}, 1966 (1972).

\bibitem{bonner} J. C. Bonner and M. E. Fisher, Phys. Rev. \textbf{135},
A640 (1964).
\end{thebibliography}
\end{document}